\documentclass[aps,twocolumn,showpacs,floats,amssymb,pra]{revtex4-1}
\usepackage{graphicx}
\usepackage{dcolumn}
\usepackage{bm}
\usepackage{color}
\usepackage{amsmath}
\usepackage{subfigure}
\usepackage{enumitem}
\usepackage{ucs}
\usepackage[utf8x]{inputenc}
\usepackage{epstopdf}
\usepackage{hyperref}
\usepackage{xcolor}
\usepackage[normalem]{ulem}

\begin{document}

\author{L. A. Pe\~{n}a Ardila$^{1,2}$, N. B. J\o rgensen$^{1}$, T. Pohl$^{1}$, S. Giorgini$^{3}$, G. M. Bruun$^{1}$ and J. J. Arlt $^{1}$}

\affiliation{$^{1}$Institut for Fysik og Astronomi, Aarhus Universitet, 8000 Aarhus C, Denmark}
\affiliation{$^{2}$Max Planck Institute for the Physics of Complex Systems, N\"othnitzer Str. 38, 01187, Dresden, Germany}
\affiliation{$^{3}$Dipartimento di Fisica, Universit\`a di Trento and CNR-INO BEC Center, I-38123 Povo, Trento, Italy}

\title{Analyzing  a Bose polaron across resonant interactions} 

\begin{abstract} 
Recently, two independent experiments reported the observation of long-lived polarons in a Bose-Einstein condensate, providing an excellent setting to study the generic scenario of a mobile  impurity interacting with a quantum reservoir. Here, we expand the experimental analysis by disentangling the effects of trap inhomogeneities and the many-body continuum in one of these experiments. This makes it possible to extract the energy of the polaron at a well-defined density as a function of the interaction strength. Comparisons with quantum Monte-Carlo as well as diagrammatic calculations show good agreement, and provide a more detailed picture of the polaron properties at stronger interactions than previously possible. Moreover, we develop a semi-classical theory for the motional dynamics and three-body loss of the polarons, which partly explains a previously unresolved discrepancy between theory and experimental observations for repulsive interactions. Finally, we utilize quantum Monte-Carlo calculations to demonstrate that the findings reported in the two experiments are consistent with each other. 
\end{abstract}

\maketitle

\section{Introduction}
\label{sec:intro}

Mobile impurities immersed in a quantum many-body environment can lead to the formation of quasiparticles. A paradigmatic example is an electron coupled to the lattice vibrations in a surrounding dielectric, which leads to the formation of a polaron~\cite{Landau48}. Polarons are central for understanding colossal magnetoresistance materials~\cite{Mannella05}, semiconductors~\cite{Gershenson06}, and have been hypothesized to be important in high-$T_\text{c}$ superconductors~\cite{Lee06}. The concept of impurity particles interacting with a surrounding medium also extends into other branches of physics, ranging from the standard model~\cite{Higgs} to atmospheric physics~\cite{atmos} . 

Progress in the field of ultracold quantum gases has allowed the study of polaron physics with unprecedented control and in novel regimes. This includes the recent series of experiments with impurities in ultracold Fermi gases, which allowed detailed studies of a quasiparticle coined the Fermi polaron~\cite{Schirotzek09, Koschorreck12, Kohstall12, Cetina16, Scazza17}. As a result, we now have a good understanding of the Fermi polaron even for strong interactions~\cite{Massignan14}. These pioneering experiments have in turn triggered studies of Rydberg impurities~\cite{Kleinbach18,Camargo18}, and renewed interest in impurity physics in solid-state systems~\cite{Sidler17}.

\begin{figure}[t!]
\begin{centering}
\includegraphics{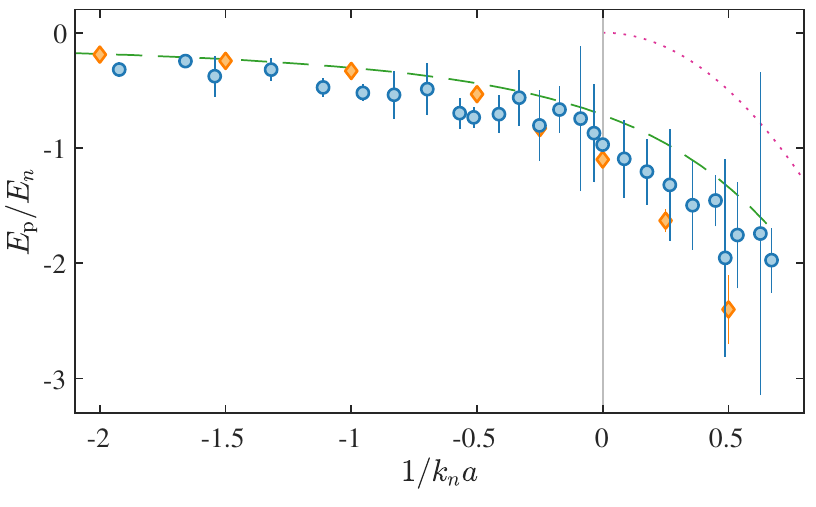}
\end{centering}
\caption{Bose polaron energy $E_\text{p}$ as a function of inverse interaction strength (blue circles), obtained from a refined analysis of experimental data from Ref.~\cite{Jorgensen16}, compared to quantum Monte-Carlo calculations (orange diamonds) and diagrammatic calculations (green dashed line). Additionally, the molecular energy (magenta dotted) is shown and the location of resonant interactions is indicated (vertical gray line).}
\label{Fig:attPolFig}
\end{figure}

Recently, the Bose polaron, i.e.\ a quasiparticle formed by a mobile impurity interacting with a surrounding Bose-Einstein condensate (BEC), was observed in two parallel experiments, at Aarhus University~\cite{Jorgensen16} and JILA~\cite{Hu16},  using different physical systems and techniques. Such Bose polarons bear stronger analogies to the generic solid-state impurity problem, since the surrounding media of both systems are bosonic, and their low-energy dispersion relations are linear. In contrast to the Fermi polaron, the BEC does not suppress interactions beyond the two-body regime, which gives rise to a broad spectrum of interesting phenomena but also renders the theoretical description challenging.

The Bose polaron has been studied  theoretically using a wide range of methods, ranging from the so-called Fr\"ohlich-Bogoliubov Hamiltonian, which is valid for weak interactions~\cite{Tempere09,Blinova13,Grusdt2015,Vlietinck15}, mean-field and variational calculations~\cite{Cucchietti06,Kalas06,Bruderer08,Li14,Levinsen15,Shchadilova16,Drescher18,Dehkharghani18,Yoshida18}, 
field theory~\cite{Rath13,Christensen15,Camacho2018}, the virial expansion~\cite{Sun17}, and Monte-Carlo 
calculations~\cite{Ardila15,Ardila16,Parisi16}. Furthermore, Bose polarons with long-range interactions have  been considered in dipolar BECs~\cite{Kain14,DipolarPolaron18}. This intense activity has improved our understanding of the Bose polaron significantly, but questions regarding its properties for strong interactions remain open.

In this work, we refine the analysis of the recent experimental results at Aarhus University~\cite{Jorgensen16} to obtain the polaron energy $E_\text{p}$ at a well-defined density across resonant interactions which is shown in Fig.~\ref{Fig:attPolFig} as a function of the inverse interaction strength. The energy is extracted from the experimental data by using a physically motivated line shape function including the many-body continuum and trap effects. This advanced analysis is the main result of our paper and provides a more detailed characterization of the Bose polaron compared to  previously reported results~\cite{Jorgensen16,Hu16}. The energies are compared to exact Quantum Monte-Carlo (QMC) and diagrammatic calculations.
 
We furthermore develop a semi-classical theory for the dynamics and decay of the polaron. This is used to shed light on a previously unresolved discrepancy between theory and experimental data for the polaron energy at repulsive interactions. Finally, our QMC calculations are shown  to also reproduce the results of the JILA experiment~\cite{Hu16}, which  demonstrates the consistency of the two experiments that have been performed under quite distinct conditions, i.e. at different gas parameters of the BEC, for different atomic species,  and using different detection methods.

The paper is organized as follows. We first present the new analysis of the attractive polaron data in Sec.~\ref{sec:attBranch}. Then, we move on to discuss the repulsive branch  in Sec.~\ref{sec:repBranch}, and introduce several effects which account for the observed discrepancy between experiment and theory. In Sec.~\ref{sec:JILA}, we perform a comparison with the observations presented in Ref.~\cite{Hu16}. Finally,  in Sec.~\ref{sec:theory}, we briefly review the theoretical methods used, and draw conclusions in Sec.~\ref{sec:conclusions}.

\section{Attractive Polaron Branch Analysis}
\label{sec:attBranch}

The energy of the attractive Bose polaron is not directly accessible from spectroscopic data due to the inhomogeneity of the trapped atomic cloud and the presence of a many-body continuum of states. Thus detailed knowledge of the line shape function is necessary to gain access to the polaron energy. In the following, such a line shape function is developed based on the relevant physical effects. We use the momentum scale $k_n=(6\pi^{2}n_\text{peak} )^{1/3}$, energy scale $E_n=\hbar^2k_n^2/2m_\text{B}$, and interaction scale $k_n a$, where $n_\text{peak}$ is the density of the condensate at the center of the trap, $m_\text{B}$ is the mass of BEC bosons, and $a$ is the impurity-boson scattering length.

Briefly, the spectroscopic investigation of the Bose polaron was performed as follows in the Aarhus experiment. Radio-frequency (rf) spectroscopy was performed on $^{39}\mathrm{K}$ impurity atoms in a $^{39}\mathrm{K}$ BEC of $n_\text{peak} = 4.0\times10^{14}$\;cm$^{-3}$ with a boson-boson scattering length between the atoms in the BEC on the order of $a_\text{B} \approx 9a_0$, where $a_0$ is the Bohr radius. If the polaron state is well defined, impurities are generated by transfer of BEC atoms into the impurity state when the rf pulse of frequency $\omega_\text{rf}$ fulfills $E_\text{p} = \hbar (\omega_0 - \omega_\text{rf} )$, where $\omega_0$ is the bare transition frequency. After the rf pulse, the atoms were released and observed after time-of-flight expansion. Due to three-body recombination between an impurity atom and two medium atoms, the impurity itself is lost and the relative fraction of lost atoms was therefore employed as spectroscopic signal.

For strong interactions, the spectral response contains a broad continuum originating from excited many-body states. Moreover, the experiment was performed in a harmonic trap such that the BEC had an inhomogeneous density distribution. The observed signal therefore contained contributions from polarons created across a distribution of densities, which resulted in a broad impurity spectrum, even when the many-body continuum was negligible. This was accounted for by comparing with theoretical calculations, which took the inhomogeneous BEC and the many-body continuum into account. Good agreement for the average energy of the impurity was obtained between experiment and theories, which assumed an infinite polaron lifetime~\cite{Jorgensen16}. Thus, long lived polarons were observed. 

To extract the polaron energy at a fixed density directly we construct  a model for the line shape of the impurity spectrum, which contains the polaron energy $E_\text{p}$ at the  peak density of the condensate as a fitting parameter. Three  relevant effects are included in the model. First, the spectral signal contains contributions from impurities created at densities ranging from zero at the edge of the condensate to $n_\text{peak}$ at the condensate center. Second, the spectral weight is extended towards higher energies due to the many-body continuum. Third, the finite duration of the rf-pulse provides a Fourier limited energy resolution.

\begin{figure}[t]
\begin{centering}
\includegraphics{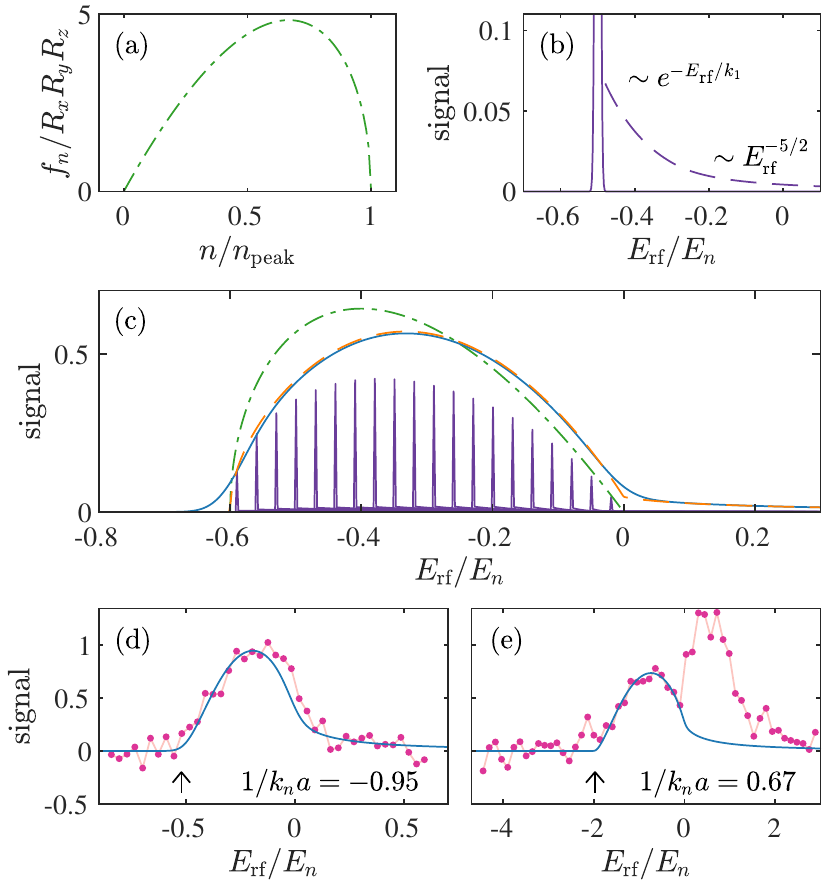}
\end{centering}
\caption{Fitting procedure to obtain the attractive polaron energy at the peak density from the spectroscopic signal of an impurity in a harmonically trapped BEC. (a)~Density distribution of the BEC in the Thomas-Fermi limit. (b)~Spectroscopic response of the impurity at a given density. The signal consists of a delta function and a tail towards higher-lying energies. (c)~The inhomogeneous BEC distribution and the signal from the impurity at a well-defined density is combined to create a line shape function. The green dash-dotted curve is the signal without the influence of the many-body continuum, and the purple curves correspond to the contributions from impurities at selected densities. Note that the many-body continuum is included, but hardly visible. The sum of contributions at all densities provides the orange dashed curve. The solid blue curve is obtained by convolution with a Gaussian corresponding to the influence of a finite rf pulse. (d-e)~The fit is applied to the experimental data of Ref.~\cite{Jorgensen16} to obtain the polaron energy (arrows) at the given interaction strengths. Note that both the attractive and the repulsive branch can be observed in panel (e). The repulsive branch is neglected in the fit.}
\label{Fig:polaronFitCartoon}
\end{figure}

The model is based on the distribution of the BEC density, $n$, in the Thomas-Fermi limit 
\begin{equation}
f_n (n)= 4 \pi R_xR_yR_z \frac{n}{n_{\text{peak}}}\sqrt{1-\frac{n}{n_{\text{peak}}}}\;,
\label{eq:dens_dist}
\end{equation}
where $R_i$ are the Thomas-Fermi radii of the condensate in direction $i=x,y,z$. This expression is shown in Fig.~\ref{Fig:polaronFitCartoon}(a) and provides the number of atoms $f_n (n) dn$ located within a density range from $n$ to $n+dn$. Since the mean-field energy of the polaron, $E_\text{MF}=2\pi \hbar^2na/m_\text{r}$ with $m_\text{r}=m_\text{B}m/(m_\text{B}+m)$ and $m$ the impurity mass, is proportional to the BEC density, it also gives the energy distribution of polarons in the weak coupling limit.

For stronger interactions, the polaron energy deviates from the mean-field value $E_\text{MF}$. Inspired by the weak coupling result Eq.~(\ref{eq:dens_dist}), we use the empirical expression
\begin{align} 
s(E_\text{rf})=A \frac{E_\text{rf}}{E_\text{p}} \sqrt{1 -  \frac{E_\text{rf}}{E_\text{p}}}
\label{eq:dens_signal}
\end{align}
to describe the spectral line shape due to the inhomogeneous density distribution, where $E_\text{rf} = \hbar (\omega_0 - \omega_\text{rf})$ is the detuning from the atomic transition frequency, and $E_\text{P}$ is a fitting parameter providing the  polaron energy in the center of the trap. The fitting parameter $A$ accounts for the overall signal amplitude.  Note that a similar expression was used to describe a spectrum associated with the Rydberg polaron~\cite{Camargo18, Schmidt_Ryd_Pol}.

We furthermore account for the spectral weight at higher energies due to excited many-body states, referred to as the many-body continuum. As shown schematically in Fig.~\ref{Fig:polaronFitCartoon}(b), an impurity has a certain fraction $Z$ of its spectral weight located at the polaron energy, and the remaining weight $1-Z$ located in the many-body continuum, where $Z$ is the quasiparticle residue. Since $Z$ depends on the density of the BEC, it changes throughout the trap: Near the edge of the condensate where interactions are weak $Z$ is approximately $1$, and it decreases with increasing interaction strength towards the center where it reaches $0 < Z_\text{peak} < 1$. In the perturbative regime, this dependence is given by  $Z^{-1}=1+Ba^2/a_B\xi$ where $B$ is a constant depending on the mass ratio $m/m_\text{B}$ and  $\xi = 1/\sqrt{8\pi na_{B}}$ is the healing length of the BEC~\cite{Christensen15}. Since $\xi^{-1}\propto\sqrt{ E_\text{MF}}$ for weak coupling, we empirically include the effects of a density dependent quasiparticle residue using

\begin{align} 
 Z(E_\text{rf})^{-1} = 1+(Z_\text{peak}^{-1} - 1) \sqrt{E_\text{rf}/E_\text{P}}.
\label{eq:Z_signal}
\end{align}

To model the spectral shape of the many-body continuum, we use a relation between the high frequency behaviour of the impurity spectral function and the contact in the BEC~\cite{Braaten10}. For large energies, the spectral function is proportional to $[(\hbar/m_\text{B}a^2 + \omega)\omega^{3/2}]^{-1}$, where $\hbar\omega$ is the energy relative to the local polaron energy~\cite{Braaten10, Wild12}. Since this power law only holds for high energies, it is necessary to introduce a function modeling the low energy behavior of the many-body continuum. Below a certain energy $\eta$ relative to the local polaron energy, we model the spectral response of the many-body continuum by an exponential $k_2 \exp(-E_\text{rf}/k_1)$. The parameters $k_1$ and $k_2$ are determined by matching the height and slope of the exponential at $\eta$ to the power-law continuum. In Fig.~\ref{Fig:polaronFitCartoon}(b), our assumption for the spectral response of an impurity at a given density is shown, including the many-body continuum. The amplitude of the full continuum is adjusted compared to the polaron peak such that its relative spectral weight is $1-Z$. The choice of $\eta$ then influences how the spectral weight is distributed within the many-body continuum, but as we shall see below, this has little effect on the obtained value for the polaron energy. 
 
In Fig.~\ref{Fig:polaronFitCartoon}(c), we show how the density distribution and the many-body continuum are combined to construct a line shape for the harmonically trapped system. The starting point is Eq.~(\ref{eq:dens_signal}) given by the dash-dotted green curve, which corresponds to the signal, if the quasiparticle residue $Z$ is unity for all densities. It yields a signal that extends from zero to the polaron energy in the center.  Now, every infinitesimal part of this signal corresponds to a different medium density and thereby a different $Z$. This is indicated in Fig.~\ref{Fig:polaronFitCartoon}(c) by the sharp peaks, each corresponding to a contribution from a local density. The height of the peaks match the green line near $E_\text{rf}=0$, whereas they are smaller for lower energies. This reflects that the residue decreases with increasing density. The spectral signal from each density including the many-body continuum is summed to provide the full line shape function indicated by the orange dashed line in Fig.~\ref{Fig:polaronFitCartoon}(c). This shows that the inclusion of the many-body continuum and a  $Z<1$  results in a signal tail at high energies as well as a  modified line shape compared to the one provided by Eq.~(\ref{eq:dens_signal}). Finally, the blue line gives the spectral shape after it  is convoluted with a Gaussian of fixed width to account for the finite length of the rf-pulse.  
  
The final line shape function thus contains fitting parameters $E_\text{p}$, $Z_\text{peak}$, and $A$. To determine $\eta$, we fit our spectral shape to the theoretical spectrum of the attractive branch provided in Ref.~\cite{Jorgensen16}, where the spectral response of the impurity in  the harmonically trapped BEC is modeled using a variational method, using various trial values of $\eta$. It turns out that the choice of $\eta$ has little influence on the obtained $E_\text{p}$ whereas $Z_\text{peak}$ varies with $\eta$. The final value of  $\eta$ is determined so that the fitted $Z_\text{peak}$ at weak interactions matches the perturbative result~\cite{Christensen15}. The fit successfully replicates the theoretical line shape and provides values of $E_\text{p}$ and $Z_\text{peak}$ at weak and intermediate interactions. At strong interactions, the fitting procedure fails to capture the theoretical line shape perfectly, but still provides an accurate measure of $E_\text{p}$. Note that the lower edge of the attractive branch polaron spectrum should always correspond to the polaron ground state at the peak density of the condensate, which implies that to obtain $E_\text{p}$ the primary task is to locate this edge, which the fit does well.
 
We apply the fitting model to the experimental data for the attractive branch of Ref.~\cite{Jorgensen16}. The spectral shape is shown for two different interaction strengths in Fig.~\ref{Fig:polaronFitCartoon}(d-e). In the fit, the repulsive branch is neglected by excluding data with positive energy at repulsive interactions $1/k_na >0$. As shown, the fit captures the experimentally observed line shape for negative energies, and the energy of a polaron located in the trap center is thereby successfully obtained. 

In Fig.~\ref{Fig:attPolFig}, we show the polaron energy obtained by applying our fitting procedure to the experimental data for the attractive branch of Ref.~\cite{Jorgensen16}. We also plot the attractive polaron energy obtained from QMC calculations and from the commonly used ladder approximation. The theoretical results are consistent with the polaron energy extracted from the experimental data, even across unitarity $1/k_na=0$. 

Our results provide a characterization of the attractive Bose polaron in the unitary regime and into the region of repulsive interactions. In particular our technique extracts the polaron energy at a well-defined density in the center of the BEC from the broad experimental spectra, by carefully taking into account the broadening effects coming from the trap inhomogeneity, the many-body continuum, and the finite pulse length. Figure~\ref{Fig:attPolFig} also demonstrates the accuracy of the QMC as well as that of  the ladder approximation. At $1/k_na \approx 0.5$, the QMC and the experimental data start to systematically deviate, which we attribute to a loss of accuracy of the fitting model under these conditions. 

At unitarity $1/k_na=0$, the observed polaron energy is also compatible with a variational theory including up to four-body correlations, provided  we use the  relation $a_{-} \approx -2 \times 10^4 a_\text{B}$ (giving $n_\text{peak}^{1/3}\left|a_{-}\right|\approx 70$)  between the three-body parameter $a_-$  and $a_B$~\cite{Yoshida18}. This linear relation between $a_-$  and $a_\text{B}$ is conjectured to hold based on QMC calculations and it is not obvious that  it  holds for the actual experiment, making the agreement between experiment and the theory presented in Ref.~\cite{Yoshida18}, which depends on the specific value of $a_-$, intriguing.

We also note that the observed spectral signal agrees with the results of Ref.~\cite{Shchadilova16}.  Interestingly this work predicts that below the observed spectrum, there is a low energy state which however is essentially unobservable since it is highly dressed by Bogoliubov modes and thus has a very small spectral weight.

It is worth noting that the large peak density of the atomic cloud implies that the local condensate fraction in the center is close to unity. Consequently, an advantage of the fitting procedure is that it provides a characterization of the polaron in an approximately pure condensate environment with a local condensate density fraction close to unity, corresponding to essentially zero temperature.

The fitting procedure is in principle capable of also providing $Z_\text{peak}$ from the shape of experimental spectrum. However, the line shape depends only weakly on $Z_\text{peak}$, and the obtained value is therefore highly susceptible to experimental noise in the spectrum. The data quality does therefore not permit to extract an accurate value for $Z_\text{peak}$. 
  
We now briefly explain why it is not straightforward to apply our fitting procedure to the repulsive branch.  The energy of the repulsive polaron is \emph{maximal} in the center of the trap, and it would therefore be tempting to determine it by locating the upper edge of the observed spectrum. However, the presence of the continuum of many-body states of arbitrarily high energy makes the upper edge of the spectrum ill defined. Thus, an accurate determination of the repulsive polaron energy using a fitting procedure similar to the one described above would therefore require a quantitatively accurate theory for the many-body continuum, which is not available except for weak coupling. Moreover, as discussed in the following subsection, the spectrum of the repulsive branch is distorted due to impurities leaving the BEC.

\section{Repulsive Polaron Branch Analysis}
\label{sec:repBranch}

In this section, we analyze the repulsive polaron spectrum observed in Ref.~\cite{Jorgensen16}. The main purpose is to discuss the discrepancy between the experimental data and theory. To do this, we develop a semi-classical model for the dynamics and three-body decay of the repulsive polaron, which is important for the Aarhus experiment, since  it uses the three-body decay of the impurities as the main observable. Moreover, the influence of the many-body continuum is estimated, which allows an approximate comparison between QMC calculations and experimental data. We find that the discrepancy can be partly explained by including these effects.

\begin{figure}[tb]
\begin{centering}
\includegraphics{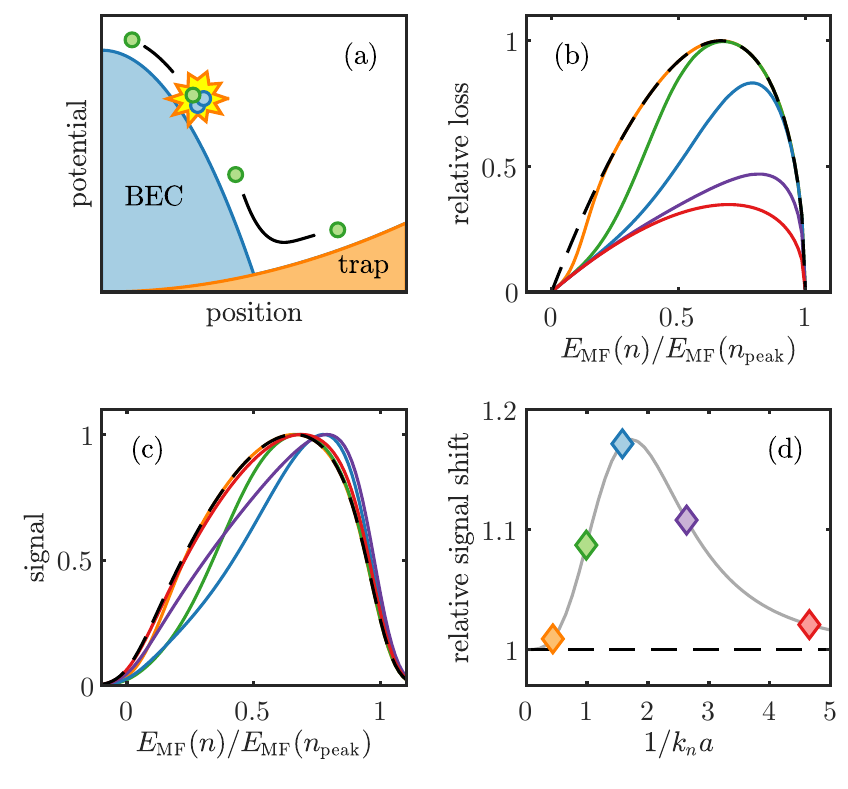}
\end{centering}
\caption{Model for impurity dynamics, which predicts a signal shift for the repulsive polaron branch. (a)~Sketch of the model. An impurity created near the center of the repulsive BEC has a large probability of undergoing three-body recombination, whereas an impurity near the perimeter has an increased probability of escaping the condensate without recombination. (b)~Predicted relative loss from the model at different interaction strengths $1/k_{n}a$ ($a/a_0$), where $k_{n}$ is calculated using the average density. From bottom to top, the red curve is $4.6$ ($170$), the purple is $2.6$ ($300$), the blue is $1.6$ ($500$), the green is $1.0$ ($800$), and the orange is $0.44$ ($1800$). The black dashed curve is the expected loss when all impurities undergo recombination. (c)~Curves of (b) converted into signal by convolution with a Gaussian function and normalization. (d)~The predicted relative signal position at different values of $1/k_na$. The shifts obtained from curves shown in (b) and (c) are shown as diamonds.}
\label{Fig:repPolFig}
\end{figure}

Consider a single impurity experiencing a repulsive mean-field potential from the BEC. Due to the interaction and the inhomogeneous density distribution of the BEC, the impurity  is accelerated outwards upon creation. Thus, the impurity has a non-zero probability of escaping the condensate  without undergoing three-body recombination, as illustrated in Fig.~\ref{Fig:repPolFig}(a). This probability is large for an impurity created in the outer low-density region of the BEC, where the recombination rate $\propto n^2$ is low and the distance to the non-condensate region small. In contrast, for an impurity created near the dense center the acceleration is lower and the recombination rate is greater, which results in a low probability for escaping the condensate. As explained above, the Aarhus experiment uses the number of atoms lost from the BEC via three-body decay as a spectroscopic signal for polaron formation. Here, it was assumed  that all impurities have the same probability of undergoing three-body recombination. However, since impurities in center of the BEC have a higher probability for undergoing three-body loss compared to those created near the edge, the loss signal for the repulsive polaron branch is skewed towards higher energy. This offers an explanation for the higher lying experimentally observed energy of the repulsive polaron branch with respect to theoretical predictions.

In the following, this effect is analyzed quantitatively. Consider a BEC in a spherically symmetric trap of frequency $\omega = (\omega_x \omega_y \omega_z)^{1/3}$, with peak density $4.0\times10^{14}$\;cm$^{-3}$, and $\omega_x$, $\omega_y$, and $\omega_z$ corresponding to those of Ref.~\cite{Jorgensen16}. A classical impurity particle of mass $m = m_\text{B}$ is accelerated by the mean-field potential $E_\text{MF}(r) = 2\pi\hbar^2an(r)/m_{r}$, where $n(r) = n_\text{peak} (1 - r^2/R^2)$ with $r$ being the radial coordinate and $R$ the Thomas-Fermi radius. The force acting on the particle is given by the gradient of $E_\text{MF}(r)$, which provides a classical equation of motion $\ddot r(t)= Cr(t)/m$, with $C = 8 \pi \hbar^2 a n_\text{peak}/mR^2$. This is the equation of an inverted harmonic oscillator potential, which for starting conditions $\dot r = 0$ and $r(0) = r_0$ has the solution $r(t) = r_0 \cosh (\sqrt{C/m} t )$.

The probability for an impurity to undergo three-body recombination can be obtained from the corresponding  recombination rate and the time it takes the impurity to leave the condensate. The three-body recombination rate is given by $\Gamma_3=L_3  n^2(r)$, where $L_3$ is the three-body recombination coefficient, with $L_3\propto a^4$ in the weak coupling regime, ignoring possible effects of Efimov resonances. An expression for $L_3$ for a single-component gas is provided by Eq.~(9b) of Ref.~\cite{Ferlaino11} (see also Ref.~\cite{Braaten06}). To calculate the probability of an impurity created at $r_0$ to undergo three-body recombination, the recombination rate is integrated over the time the impurity takes to leave the condensate, i.e. 
\begin{align} 
\zeta(r_0) \equiv \int_0^{t_\text{out}} L_3  n^2[r(t)] dt,
\end{align} 
where $t_\text{out} = \sqrt{m/C} \;\text{acosh} (R/r_0)$, and $r(t) = r_0 \cosh (\sqrt{C/m} t )$. This expression describes the average number of decay events the impurity will undergo traveling from $r_0$ to the edge of the BEC.   It can be converted into a probability for undergoing three-body decay using 
\begin{align}
P_\text{3BR} = 1 - e^{-\zeta}.
\end{align}
For an impurity created at a specific position $r_0$ and an interaction strength $a$, it is thus possible to calculate the probability of  recombination before leaving the condensate.

Towards strong coupling, we no longer have $L_3\propto a^4$. To include this effect in the model, we replace the scattering length $a$ by an effective length $a_\text{eff}^{-1} = a^{-1} + n^{1/3}$ so that  $L_3\propto a_\text{eff}^4$, which qualitatively captures the transition to the unitary regime $1/a = 0$, where the inteparticle spacing $n^{-1/3}$ is the relevant interaction length scale~\cite{Makotyn14, Eigen17}.

 If an impurity undergoes three-body recombination, three atoms are lost, but if the impurity instead leaves the condensate, only the one impurity atom is lost. Thus, the net atom loss due to an impurity created at position $r_0$ is 
 \begin{align}
s_\text{loss}(r_0)= 1+2P_\text{3BR}(r_0)
\label{eq:sloss}
 \end{align}

Based on these arguments it is possible to examine the impact on the observed spectral shape. All equations presented above are solved numerically with an effective scattering length $a_\text{eff}$. Figure~\ref{Fig:repPolFig}(b) shows $f_n s_\text{loss}$ obtained from Eqs.~\eqref{eq:dens_dist} and \eqref{eq:sloss} as a function of the corresponding mean-field shift $E_\text{MF}=2\pi \hbar^2na/m_r$ for various coupling strengths. Since a given position $r_0$ corresponds to a specific density and mean-field energy, the loss $s_\text{loss}$ can be regarded as a function of  $E_\text{MF}$. This plot gives the mean-field spectral line shape including  a spatially dependent three-body loss. For comparison, the line shape 3$f_n$ assuming all impurities undergo three-body recombination is also shown. The curves have been scaled such that the peak value of $f_n s_\text{loss}$ is 1. A spatially dependent loss rate significantly affects the spectral shape. The signal at high energy is affected the least, since it originates from impurities created near the center, which have a high chance of undergoing three-body loss. On the other hand, the signal at low energies due to atoms created nearer the edge is strongly reduced by the significant probability of escaping the BEC without undergoing three-body loss. Note that for strong coupling all impurities undergo three-body loss and the spectral shape is simply $3f_n$, whereas for weak coupling almost no atoms undergo three-body loss and the spectral shape is  approximately given by  $f_n$. It follows that the spectral \emph{shape} deviates most from the functional form   $f_n$ for \emph{intermediate} coupling strengths. As we shall see, this non-trivial conclusion turns out to be confirmed when comparing with the experimental results. 

To allow for comparison with the experimentally obtained signal, the line shape obtained using the model above is convoluted with a Gaussian of width $\sigma = E(n_\text{peak})/15$, and normalized as shown in Fig.~\ref{Fig:repPolFig}(c). For simplicity, we have chosen Gaussians with widths, which are constant relative to $E_\text{MF}$.

\begin{figure}[t]
\begin{centering}
\includegraphics{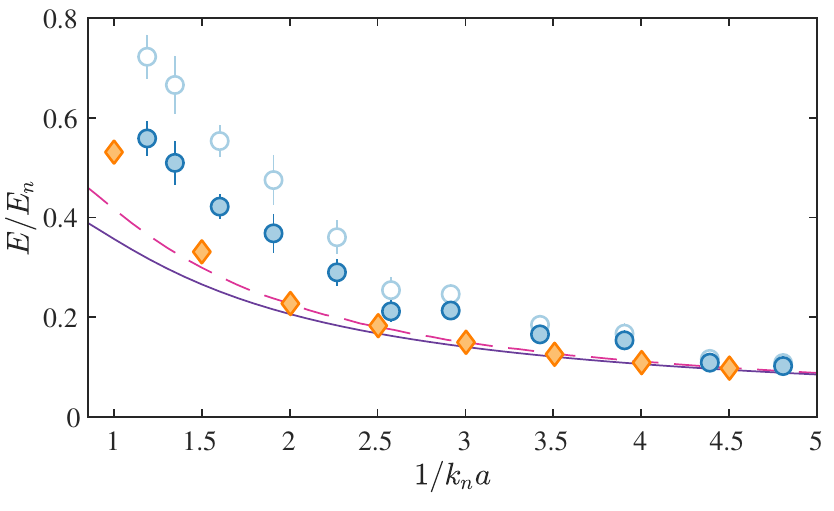}
\end{centering}
\caption{Repulsive polaron branch analysis. The purple line is obtained from a variational truncated basis method approach using the average density. The magenta dashed line is also obtained using the truncated basis method, but shows instead the average energy of the theoretical spectrum including an inhomogeneous medium and the many-body continuum. The orange diamonds are obtained from QMC calculations using the average density. The blue data points are experimentally obtained energies. The light blue data points show the average energy of the experimental spectrum. The filled blue data points have been scaled to compensate for impurity dynamics and decay, and the many-body continuum.}
\label{Fig:repulsiveBranch}
\end{figure}

In the previous evaluation of the experimental results, an average energy of the polaron spectrum was obtained by fitting a Gaussian to the observed signal~\cite{Jorgensen16}. To check if our present model can explain the observed discrepancy between theory and experiment, we fit a Gaussian to the line shapes described above. The obtained average energies are shown in Fig.~\ref{Fig:repPolFig}(d) normalized to the energy obtained from a fit to the curve assuming all impurities undergo recombination. Impurities leaving the condensate without undergoing three-body loss clearly give a significant relative shift in the obtained energy. Remarkably, the shape of the relative shift is very similar to the observed discrepancy between experiment and theory, see Ref.~\cite{Jorgensen16} and Fig.~\ref{Fig:repulsiveBranch}. In particular, there is a significant upward shift in the average energy for intermediate coupling strengths, which is of similar magnitude as the observed discrepancy. This suggests that we have indeed identified at least part of the reason for the observed upward shift in the average energy as compared to theories for the repulsive polaron: It is due to the fact that impurities created near the center of the BEC, have a larger probability for undergoing three-body decay compared to impurities created near the edge. 

To examine this further, Fig.~\ref{Fig:repulsiveBranch} shows experimental as well as theoretical data for the repulsive polaron branch. The light blue  points show the experimental data  of Ref.~\cite{Jorgensen16} for the average energy of the observed spectrum. The filled blue points are obtained by scaling this data using the model above to account for the three-body loss dynamics. In addition, the many-body continuum above the repulsive polaron energy has been taken into account. This is done to compare with QMC calculations, which do not include the many-body continuum. A scaling factor accounting for the many-body continuum is estimated by taking the ratio between the average energy obtained from a variational calculation~\cite{Jorgensen16} with (magenta dashed line) and without (purple line) the many-body continuum. When these effects are taken into account, the experimental data (full blue points) are lowered compared to the average observed energy (light blue points). As a result, there is now indeed a better agreement with the theoretical calculations also shown in Fig.~\ref{Fig:repulsiveBranch}~\footnote{Calculations were also performed using $a$ instead of $a_\text{eff}$, varying $L_3$ by a constant factor, different Gaussian convolution widths, and with a signal evaluation based on the first moment of the signal distribution. All these variations provide the similar results as presented in Fig.~\ref{Fig:repPolFig}(d), with minor shifts to the location and amplitude of maximum shift. The presented results are thus robust against the exact choice of parameters in the model.}.
 
However, Fig.~\ref{Fig:repulsiveBranch} shows that there is still a significant disagreement between experimental data and theory, even after these effects are included. This could be due to several reasons. First, our semi-classical model for the impurity dynamics and recombination still involves approximations.  A more sophisticated calculation requires a comprehensive model for effective polaron energy, impurity recombination, and dynamics. Nevertheless, we believe our model provides a qualitatively correct estimate of the impact of impurity dynamics on the observed spectral signal obtained via loss measurements.

Unlike the case for the attractive polaron branch, the analysis of the repulsive branch is not restricted to the center of the Bose-Einstein condensate, where the local condensate fraction is close to unity. Therefore, the data shown for the repulsive polaron branch is more susceptible to effects from finite temperature which have been predicted to influence the spectrum~\cite{Levinsen17, Guenther18}.

Another mechanism, which is not included in our model, is the two-body decay of repulsive polarons into the attractive branch. This decay  has been shown to have a rate similar to the three-body decay for the Fermi polaron~\cite{Massignan2011}, and it also has a larger probability of occurring at greater medium densities near the trap center. It  will release energy into the system, which excites some atoms out of the  condensate and is therefore expected to influence the spectroscopy signal. 

Finally,  the theories for the repulsive polaron are approximate. This holds even for the QMC calculations, since one has to include a node in the wave function to approximately ensure orthogonality with lower lying states, when calculating the energy of the repulsive polaron. As a result, the QMC calculations are of the fixed-node type and therefore not exact. Also, our estimate for the effects of the many-body continuum is based on approximate variational calculations. 

The motional effects discussed in this section have no relevance for the spectroscopy of the attractive branch. Since impurities are  attracted to the center of the BEC for attractive interactions,  it is safe to assume that they all eventually undergo three-body decay.

\section{Comparison with Observations in a K-Rb Mixture}
\label{sec:JILA}

We now analyze the recent observation of Bose polarons in a K-Rb mixture at JILA~\cite{Hu16}. In this study, fermionic $^{40}\mathrm{K}$ atoms were employed as impurities in a $^{87}\mathrm{Rb}$ BEC with a boson-boson scattering length $a_\text{B}=100a_{0}$, and mass ratio in $m_\text{B}/m\approx2.2$. Spatially resolved imaging allowed the impurities in the approximate center of the BEC to be probed selectively resulting in an average probed density of $1.53\times10^{14}$\;cm$^{-3}$.

\begin{figure}[tb]
\begin{centering}
\includegraphics{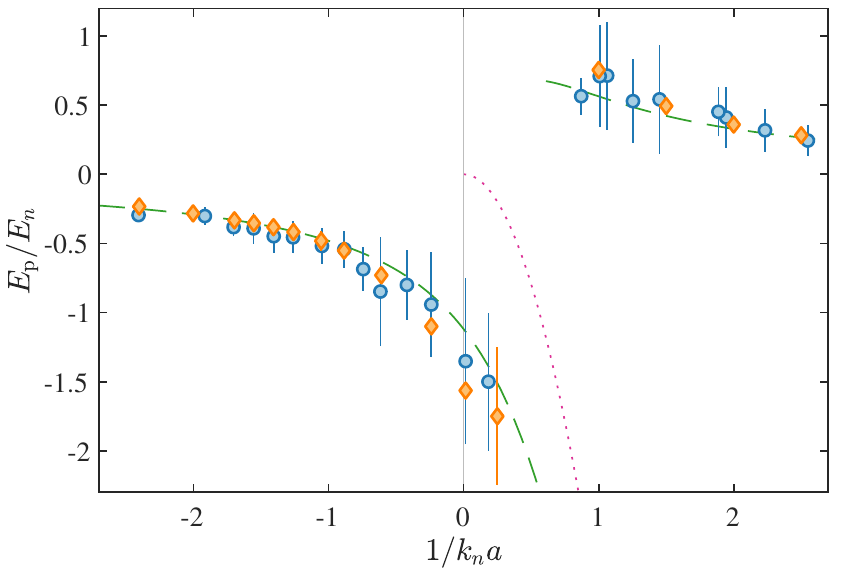}
\end{centering}
\caption{Comparison of experimental data from Ref.~\cite{Hu16} (blue circles) with quantum Monte-Carlo calculations (orange diamonds) and diagrammatic calculations (green dashed line). Additionally, the molecular energy (magenta dotted) is shown and the location of resonant interactions is indicated (vertical gray line). The error bars of the experimental data corresponds to the rms spectral width.}
\label{JILA}
\end{figure}

The main results of the experiment is shown in Fig.~\ref{JILA}. The spectrum displays two branches corresponding to the attractive and repulsive polaron at negative and positive  energies respectively. For the attractive branch, a polaron signal is detected slightly beyond unitarity, whereas for the repulsive branch, the signal is lost at $1/k_{n}a\sim1$, where the repulsive polaron becomes ill-defined.

For theoretical comparison, we show new QMC simulations for both branches, as well as results obtained from diagrammatic calculations. Both theoretical methods display agreement with the experimentally obtained results which demonstrates the consistency between the experimental results obtained in Aarhus and at JILA. 

\section{Theoretical Methods}
\label{sec:theory}

In this section, we briefly introduce the theoretical methods we use to analyze the Bose polaron, i.e the QMC and diagrammatic methods.

In the QMC calculations, we consider a single impurity of mass $m$ immersed in a gas of $N$ bosons with mass $m_\text{B}$. The Hamiltonian of the system is 
\begin{equation}
\mathcal{H}=-\sum_{i=1}^N \frac{\hbar^2\nabla_i^2}{2m_\text{B}}+\sum_{i<j}V_\text{B}(r_{ij}) 
-\frac{\hbar^2\nabla_\alpha^2}{2m}+\sum_{i=1}^N V(r_{i\alpha}).
\label{eq:Hamiltonian}
\end{equation}
 The Bose gas is taken to be weakly interacting with $n_\text{peak}a_\text{B}^{3}\ll1$. In the calculations, the boson-boson interaction potential $V_\text{B}(r_{ij})$ is modeled as a hard (impenetrable) sphere with a range, given by $a_\text{B}$, much smaller than the average interparticle distance.  The boson-impurity interaction  $V(r)$ is instead modeled by a square well or equivalently, by a zero-range pseudo-potential~\cite{Ardila16}. We solve the two-body problem to find the relation between the model potential  parameters and the impurity-boson scattering length $a$, which  can take any value as is the case in the two experiments~\cite{Jorgensen16,Hu16}. 

The initial trial  wave function serving as a starting point for the QMC simulations is optimised using the 
variational Monte-Carlo (VMC) technique for the different experimental conditions. We obtain the polaron energy $E_\text{p}$  by computing the ground-state energy of $N$ bosons and one impurity, and subtracting the energy of the system without impurity, i.e $E_\text{p}=E^{0}(N+1)-E^{0}(N)$. Within this work, the calculations have been performed with $N=64$ bosons, and size effects have been checked. The QMC method is exact for the ground-state energy within statistical errors, and it includes \textit{all} correlations for any coupling strength. The details of the QMC approach are given in Refs.~\cite{Ardila15, Ardila16}. Note that previously QMC calculations were compared with experiments using the average density of the BEC in the simulations. Here, we use the peak density of the BEC in the QMC calculations since this is what is extracted in the data analysis. In addition, we perform QMC calculations for a range of interaction strengths from weak coupling all the way across the resonance along the attractive branch, as well as along the repulsive branch on the positive-$a$ side of the resonance.
 
We also use a diagrammatic scheme to calculate the polaron energy. While this method is approximate, it complements the exact QMC calculations because it can describe the full spectral function. The details of the diagrammatic approach  are described in Refs.~\cite{Rath13,Guenther18,Camacho2018}.

\section{Conclusions}
\label{sec:conclusions}

In summary, we have presented an extended and refined analysis of the Aarhus experiment probing impurities in a BEC. For the attractive polaron, we developed a physically motivated  model for the spectral line shape to account for the effects of the trap inhomogeneity and the many-body continuum. This allowed us to extract the energy of the attractive polaron at a well-defined density from the experimental data. We compared this energy with QMC as well as diagrammatic calculations and obtained good agreement, even in the strongly interacting regime. For repulsive interactions, we developed a semi-classical theory for the dynamics and three-body loss of the polaron, which permits to partly explain the observed disagreement between theory and experiment.\\

Our new analysis  paves the way for several intriguing new research directions, including experimental studies of the interactions between polarons~\cite{Dehkharghani18,Camacho2018}
 and the formation of bipolarons~\cite{Camacho2018b}, explorations of temperature effects and associated new quasiparticle states~\cite{Guenther18}, as well as investigations of non-equilibrium effects and the formation dynamics of the polaron~\cite{Nielsen18,Shchadilova16,Lausch2018,Lampo2018}. 

\begin{acknowledgments}
We gratefully thank Arturo Camacho-Guardian for providing the diagrammatic calculations. This work has been supported by the Villum Foundation, the Carlsberg foundation, the Danish council for Independent Research. In addition,  T. P acknowledges the Danish National Research Foundation through a Niels Bohr Professorship. S. G. acknowledges support from the EU Horizon 2020 research and innovation programme under grant agreement No. 641122 QUIC, and from the Provincia Autonoma di Trento.
\end{acknowledgments}

\bibliography{BosePolExp}

\end{document}